\documentclass[twocolumn,showpacs,preprintnumbers,amsmath,amssymb]{revtex4}
\usepackage{graphicx}
\begin{document}
\title{1/f noise in nanowires} 
\author{Aveek Bid\footnote[1]{Electronic mail: avik@physics.iisc.ernet.in}$^1$, Achyut Bora$^1$ and A. K. Raychaudhuri\footnote[2]{Electronic mail: arup@bose.res.in}$^{1,2}$}        
\address{$^1$Department of Physics, Indian Institute of Science,  Bangalore 560 012,  India\\
$^2$Unit for Nanoscience and Technology, S.N.Bose National Centre for Basic Sciences, Salt Lake, Kolkata 700098, India} 

\begin{abstract}

We have measured the low-frequency resistance fluctuations (1 mHz$<$f$<$10 Hz) in Ag nanowires of  diameter 15 nm $\leq$ d $\leq$ 200 nm at room temperatures. The power spectral density (PSD) of the fluctuations has a $1/f^{\alpha}$ character as seen in metallic films and wires of larger dimension.  Additionally, the PSD   has a significant low-frequency component and the value of $\alpha$ increases from the usual 1 to $\simeq 3/2$ as the diameter d is reduced.  The value of the  normalized fluctuation $\frac{<\Delta R^2>}{R^2}$ also increases as the diameter d is reduced. We observe that there are new features in the  1/f noise as the size of the wire is reduced and they become more prominent as the diameter of the wires approaches 15nm.  It is important to investigate  the origin of the new behavior as 1/f noise  may  become a limiting factor in the use of  metal wires of nanometer dimensions as interconnects.
 
\end{abstract}
\maketitle
\section{Introduction}
Noise and fluctuation in general, are enhanced when the system size is reduced. In the case of equilibrium thermal fluctuation of a physical quantity, it is inversely proportional to the system size. In this paper, we would like to investigate the specific issue of electrical noise in metallic nanowires. This is an important issue not only as a basic  question but also as an important parameter that should  be considered for the use of  nanowires as interconnects in nanoelectronics. The equilibrium white thermal  noise (the Nyquist Noise) of a wire of resistance R  at a temperature T can be estimated from the power spectral density $S_{th}\approx 4k_BTR$ ~\cite{nyquist} where $k_B$ is the Boltzmann constant. This component of the electrical noise depends only on the sample resistance R.  The Nyquist noise has no explicit dependence on of the size of the sample, the only size dependence in through the size dependence of R. However, in a current carrying nanowire a larger contribution to the electrical noise is expected to arise from the $\lq\lq$excess noise" or $\lq\lq$conductance or resistance noise". This noise has a  spectral power density of the type $S(f)\propto 1/f^{\alpha}$, where  $\alpha \sim 1$ and is known as   $\lq\lq$1/f noise."  We note that {\it apriori} there is no  way of estimating the actual value of the 1/f noise because 1/f can arise from many different sources ~\cite{weissmann,Dutta}.  As a result, an experimental measurement is an absolute necessity to make even an estimate of this noise component. For metallic films, past studies have shown that an estimate of the value of the  1/f noise can be made from the empirical Hooge's formula  given as:
\begin{equation}
S_V(f) = \frac{\gamma_HV^2}{N.f}
\label{Hooge}
\end{equation}
where N is the number of electrons in the sample, V is the bias used for measurements and $\gamma_H$ is an empirical constant which for metallic films lie in the range  $\simeq 10^{-3}-10^{-5}$. In this paper we address the issue of  1/f electrical noise in nanowires. We have measured the low-frequency resistance fluctuations (1 mHz$<$f$<$10 Hz) in Ag nanowires of  diameter 15 nm $\leq$ d $\leq$ 200 nm at room temperatures. The wires were electrochemically grown in templates of anodic alumina or polycarbonate. We observe that the noise indeed has a spectral power density $S(f)\propto 1/f^{\alpha}$ and the magnitude of noise increases as the size is reduced. The observed noise in the metallic nanowires is larger ($ \gamma_H\sim 4 \times 10^{-3}$) than what one would expect from a simple extrapolation of Hooge's relation to this domain. The exponent $\alpha$  also deviates from $~1$ as the diameter (d) of the nanowire  is reduced.  To our knowledge, this is the first report of the investigation of 1/f noise in nanowires in this range of diameters.
\section{Experimental techniques}
The Ag nanowires have been grown within polycarbonate (PC) templates (etched track membranes) or Anodic Alumina (AA) templates   using electrochemical deposition from AgNO$_3$ solution ~\cite{akr}. The membranes have been procured from a standard commercial source  ~\cite{whatman}.  During the growth one side of the membrane was covered with an evaporated metal and was used as the anode. The other electrode was a micro-tip (radius of curvature $\simeq 100  \mu$m) that can be  positioned by a micropositioner to a specific area on the other side of the membrane and the growth can thus be localized. The arrangement is shown as inset in figure~2. The growth current used is approximately 10 mA. Generally the growth occurs within the pore by filling it from end to end. As soon as a few wires grow and touch the electrodes, the electrochemical cell gets short circuited and  the electrochemical deposition stops.  The membrane containing the nanowire is then cleaned with deionized water and annealed at 400 K for 10-12 hours. During annealing a current ~1 mA of is passed through the sample. The post deposition annealing is needed to stabilize the wire. The wires after growth can be retained within the membranes, as was done for the electrical measurements or  can be freed from the membrane by etching the membrane off with a suitable etchant (Dichloromethane for PC and KOH for AA). 

The wires  after  growth were characterized by X-ray, SEM and TEM.  Figure~1(a) shows a SEM image of the assembly of 100 nm wires (after partially dissolving off the template).  A TEM image of a 15 nm wire (taken after removing the membrane) is shown in figure~1(b).  From the TEM data we find that the wires are single crystalline with occasional twinning. The selective area diffraction patterns showed that the wires have FCC structure. The diameter of the wire is close (within $~10\%$) to the nominal diameter of the pore of the membrane. This shows that the wires grow by filling the membranes. 

\begin{figure}
\begin{center}
\includegraphics[height=5cm,width=6cm]{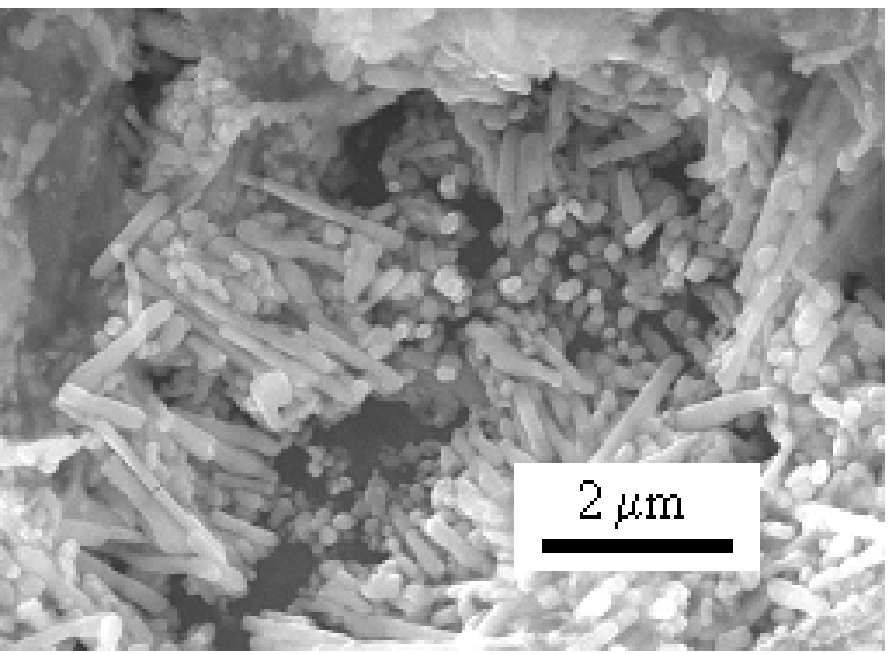} \hspace{0.5cm} \includegraphics[height=5cm,width=6cm]{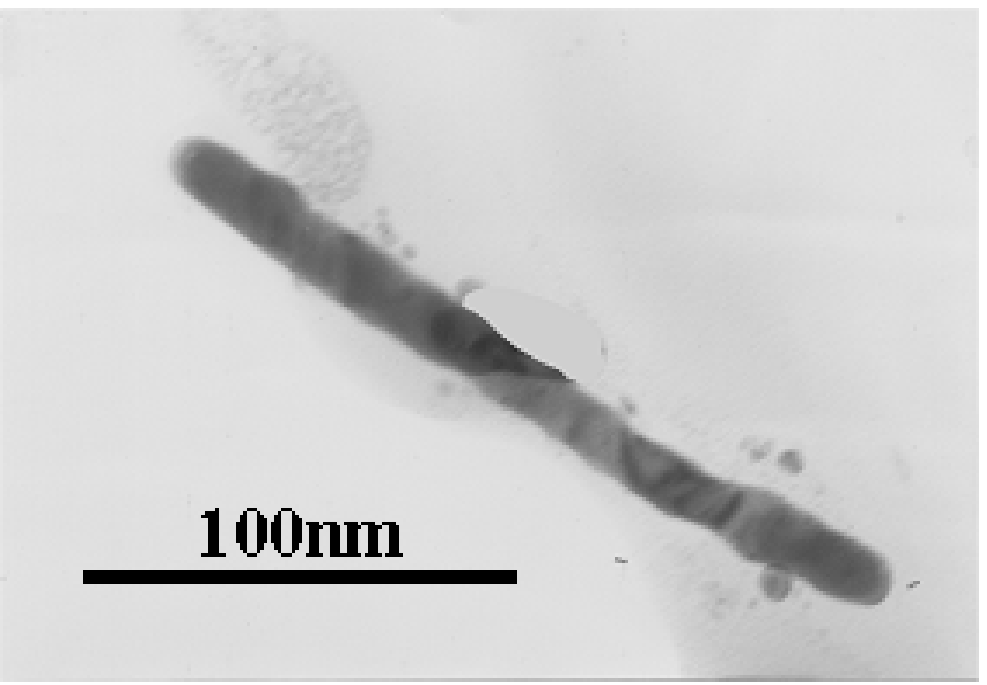}
\end{center}
\caption{\label{fig:figure1} (a)SEM image of 100 nm wires taken after dissolving the template.  (b)TEM image of a 15 nm Ag wire.}
\end{figure}

Table~1 gives the diameter and lengths of different samples that have been used in the measurements. 
\begin{table}\begin{center}
\caption{\label{tab:table1}The radius and diameters of the various samples used in this work. }
\begin{tabular}{|c|c|c|c|}
Sample&Diameter of pore &Length\\
&in membrane & of wire\\
\hline
1& 15 nm & 6 $\mu$m\\
2& 20 nm &60 $\mu$m\\
3& 30 nm &6 $\mu$m\\
4& 50 nm &6 $\mu$m\\
5& 100 nm &60 $\mu$m\\
6& 200 nm &6 $\mu$m\\
\end{tabular}
\end{center}
\end{table}
In figure~2 we show the typical XRD data for a 20 nm wire. The XRD data on the nanowire matches nicely with the data taken on a bulk silver sample. The observed lines can all be indexed to a FCC Ag.
\begin{figure}
\begin{center}
\includegraphics[height=8cm,width=10cm]{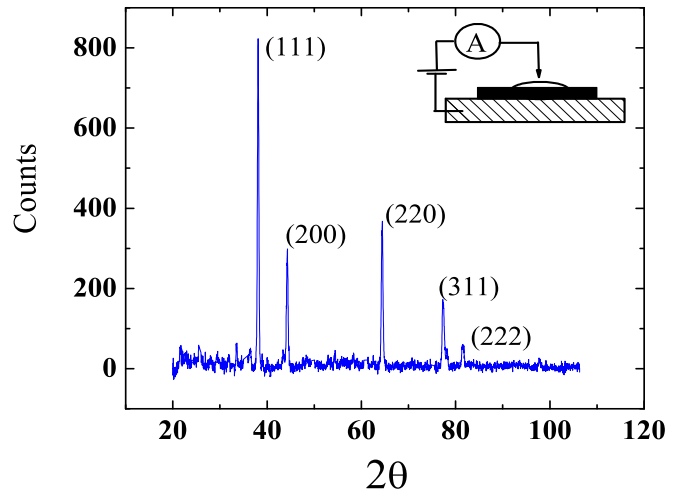}
\end{center}
\vspace{-1cm}
\caption{\label{fig:figure2} The XRD data for a 20 nm Ag wire. The XRD data on the nanowire matches with the data taken on a bulk silver sample and the observed lines can all be indexed to a FCC Ag. }
\end{figure}
We have  measured resistance of the arrays of nanowires down to 4.2 K. This was done to check the residual resistivity ratio and the behavior of the resistance at low temperatures.  The resistance noise measurements were done at room temperature using a digital signal processing (DSP) based a.c technique which allows simultaneous measurement of the background noise as well as the bias dependent noise from the sample ~\cite{scoff1,ag1}. The sample is current biased and the resistance noise appears as a voltage fluctuation. The data are taken by stabilizing the temperature with $\Delta T/T \simeq 4 \times 10^{-3} \%$. The measured background noise (bias independent) was white and was contributed by the 4k$_B$TR Nyquist noise. The apparatus can measure a spectral power down to $10^{-20}$  V$^{2}$Hz$^{-1}$.  We have used a transformer preamplifier SR554 to couple the sample to the lock-in-amplifier. The basic electrical schematic is shown in the inset of figure~3. The details of the data acquisition and the signal processing are given elsewhere~\cite{scoff1,ag1}. A single set of data are acquired typically for a time period of about fifty minutes or more at a sampling rate of 1024 points/sec. The time series of voltage fluctuations as a function of time consisting of nearly 3 million points was decimated to about  0.1 million points before the  power spectral density S$_{V}$(f) is  determined numerically. The frequency range probed by us ranges from 1 mHz to 10  Hz. The frequency range is determined mostly by practical considerations. The lower frequency limit is determined by the quality of the temperature control. The sample resistance and the bridge output may  also show a long time drift (in general such a long time drift is subtracted out by a least square fit to the data). Taking these factors into consideration the lower spectral limit in our experiment has been kept at 1 mHz. The high frequency  limit of 10 Hz is decided by the spectral window of the lock-in amplifier output filter which is operated at 24 dB/octave with a time constant of 3 ms.

The resistance and noise measurements were carried out by retaining the wires within the polymeric or alumina membranes. The sample, as stated before, consists of an array of nanowires. On each of the two sides of the membrane two electrical leads were attached using silver epoxy.  Though the measurements were made with the wires retained within the membrane, the system is an array of parallel nanowires where the individual wires are well separated by the insulating membrane. A very important issue in this measurement is the contribution of the contacts to the measurement.  We have paid attention to this aspect and measured both resistance and noise by making the contact in different ways. We discuss this issue in more detail at a later stage when we discuss the data. We find that the contact contribution, if any, is negligibly small both in the measured resistance as  well as the noise. The resistance measurements were done using the same electrodes which were used for noise measurements. 

\section{Results and Discussion}
In figure~3 we show the temperature dependence of the resistance of  Ag  nanowire arrays  of diameter 15 nm. The resistivity data for the wires of  other diameters are qualitatively similar  and are therefore not shown. The arrays contain different number of wires. It is possible to make a rough estimate of wires by a procedure described before ~\cite{giordano} to within a factor of 2.  Due to the uncertainty in the number of wires in the arrays we cannot fix the exact resistivities of the nanowires even if their diameter and the length are known to a  reasonable degree of certainty. For all the samples the resistance has a fairly linear temperature dependence  down to 75K and $R$ reaches a residual value below 30K. The  residual resistivity  ratio (RRR)$\frac{\rho_{300K}}{\rho_{4.2K}} $  for all the wires lie between 3 and 6. We find that the wires do not show any upturn in resistivity at low temperatures. This shows that the conduction of electrons in these wires are not controlled by disorder induced effects like localization~\cite{bergmann}. This ensures that the product $k_{F}l_{e}$ is much larger than 1. (Here $k_{F}$ is the Fermi wavevector and $l_{e}$ is the elastic mean free path.)  The temperature coefficient of resistivity ($\beta = \frac{1}{R} \frac{dR}{dT}$) can be obtained from the resistance data  for the nanowires. $\beta$ is in the range $\approx 2.5-3.5 \times 10^{-4}$ /K. This is smaller than but comparable to that of bulk silver which is $\approx 3.8\times 10^{-4}$ /K.  The temperature dependence of the resistivity thus shows predominantly metallic behaviour which is in agreement with the TEM observation that the wires are single crystalline.   
\begin{figure}
\begin{center}
\includegraphics[height=7cm,width=9cm]{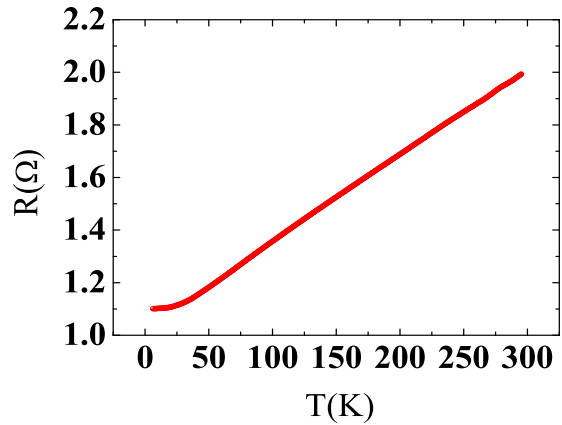}
\end{center}
\vspace{-1cm}
\caption{\label{fig:figure3} The resistance of Ag wires of 15nm diameter as a function of temperature. The inset shows a schematic of the measurement electronics.}
\end{figure}
For bulk silver at room temperature the mean free path is $\approx$ 60 nm. For the wires with diameter $<$ 50 nm, the mean free path  will have significant contribution from the surface as well as from the grain boundaries within the wire ~\cite{FS,mayadas}. This is an issue of considerable current interest due to diminishing size of the metallic interconnects in IC ~\cite{size1, size2}. Given the scope of the paper we do not elaborate on this particular issue in detail but note that these factors which limit the electron mean free path can also  contribute to the resistance noise. 

In figure~4 we show the typical low frequency power spectrum of the voltage noise arising from the resistance fluctuation in the current biased nanowires. The spectral power S$_{V}$(f)  is $\propto$ to V$^2$ where V is the bias voltage (typically few tens of $\mu$ V) and in figure~4 the spectral power is shown as S$_{V}$(f)/V$^{2}$.  The power spectra have a predominant 1/f$^{\alpha}$ nature. The observed noise appears to be independent on the nature of the membrane (polymer or alumina) as the observed noise data in wires of similar diameter are qualitatively the same.  We note that contact can be a serious issue in noise measurements and we carried out a number of checks to rule out any predominant contribution of the contact to the measured noise. In addition to silver epoxy contacts we also used  evaporated silver films and Pb-Sn solder to make contact.  We find that they give similar results to within $\pm 15\%$. In case of the Pb-Sn contact the change in R of the sample as we go below the superconducting transition temperature ($\sim$ 7 K) of the solder is negligibly small ($<2-3\%$) implying a very small contribution of the contact to the total R measured.  We have also measured the noise in different samples with varying numbers  of wires in them. The resistance of the array, due to different number of wires in them vary.  Also such samples will have significantly different contact areas and hence different contact noise, if any. Yet we find that the normalized noise $\frac{<(\Delta R)^2>}{R^2}$ in different arrays of the same diameter wire lie within $\pm 15\%$.  All these tests rule out any predominant contribution from the contacts. 
\begin{figure}
\begin{center}
\includegraphics[height=6cm,width=8.5cm]{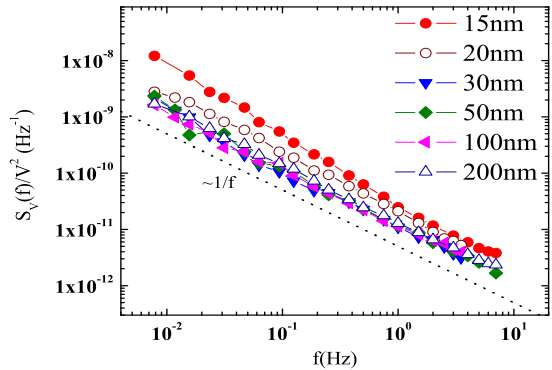}
\end{center}
\vspace{-1cm}
\caption{\label{fig:figure4} Low frequency power spectrum of the voltage noise arising from the resistance fluctuation in the current biased nanowires. The dotted line is a plot for which $\alpha =1$. }
\end{figure}

The normalized resistance fluctuation  $\frac{<\Delta R^2>}{R^2}$ for a nanowire array of resistance R is obtained by integrating  the  power spectral density $S_{V}(f)/V^{2}$ over the band width of our measurement. This equation, while giving correctly the relative resistance fluctuation of  a single wire, is also applicable when there is more than one wire in parallel because of the following reasoning. If  R the is resistance of {\bf n} identical parallel wires each of resistance R$_1$, then $1/R=n/R_1$. From this we get $\frac{<\Delta R^2>}{R^2}$ = $\frac{<\Delta R_1^2>}{R_1^2}$ thus implying  that the measured noise is independent of the number of wires in parallel.  

We make the interesting observation that resistance fluctuation has a dependence on the diameter of the wire. For this purpose we separate the wires into two groups according to their length. Group A contains wires grown in the PC membranes and have a length of 6 $\mu$ m. Group B contains wires of length 60 $\mu$ m that were grown in Alumina  templates. Thus a nanowire array belonging to a given group differ only in their diameters. The data are shown in figure~5. One can clearly see that for nanowires of a given length (group A or B) there is an increase in the normalized noise as the diameter d is reduced. 

\begin{figure}
\begin{center}
\includegraphics[height=8cm,width=10cm]{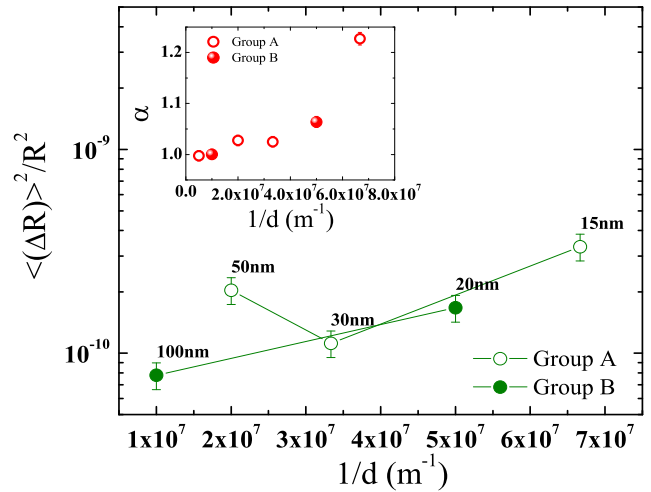}
\end{center}
\vspace{-1cm}
\caption{\label{fig:figure5} The normalized resistance fluctuation  $\frac{<\Delta R^2>}{R^2}$ at 298 K as a function of the inverse wire diameter 1/d. The inset shows the variation of $\alpha$ with 1/d.}
\end{figure}

The exponent $\alpha$ has an interesting dependence on the size of the wire as well. This is shown  in the inset of figure~5 where $\alpha$ is plotted as a function of 1/d. For  wires with diameter d $\geq$ 100 nm  the exponent $\alpha$ is $\approx$ 1. The value of the exponent is similar to what one finds in  wires of larger diameter and in metallic films. This is generally expected to arise from activated defect dynamics as envisioned in the Dutta-Horn model ~\cite{Dutta}.   However, there is a perceptible increase in $\alpha$ once the wire diameter reaches below 50nm and  it increases further when the diameter is reduced to 15 nm. Such a change in $\alpha$ is rather interesting and would indicate the presence of additional noise mechanism in the nanowires which are not present in wires of larger diameter. We are currently investigating the origin of such an extra component as well as the physical reason of the enhancement of the normalized noise on reducing the diameter d of the nanowire.

To check that such a size dependence of the 1/f noise is indeed observed in other metallic nanowires we have also carried out noise measurements in Cu nanowires of diameter 15nm grown by similar electrochemical method in PC membranes. The spectral power density is  shown in figure~6. The behavior of the spectral power is rather similar to the Ag nanowire of the same diameter and the magnitude of the resistance fluctuation $\frac{<\Delta R^2>}{R^2} \approx 1\times10^{-9} $  at 300 K which is comparable to that seen in 15 nm Ag nanowire. This strongly suggests  that the observed behavior may  be a general behavior of the metallic nanowires.

\begin{figure}
\begin{center}
\includegraphics[height=8cm,width=10cm]{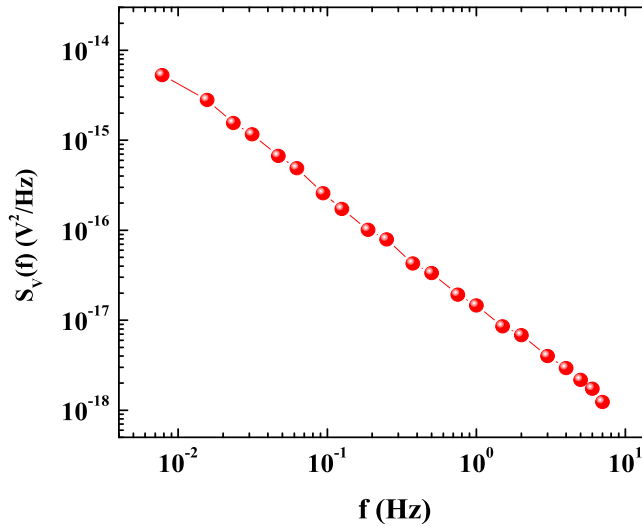}[h]
\end{center}
\vspace{-1cm}
\caption{\label{fig:figure6} Relative resistance fluctuations in Cu nanowires of diameter 15 nm grown by electrochemical method in PC membranes.}
\end{figure}

It is tempting to express the observed results in the form of Hooge relation (equation~\ref{Hooge}). However, due to the uncertainty in the exact numbers of wires  in the arrays we cannot estimate $\gamma_{H}$. Also the Hooge relation is valid for $\alpha =1$  and any  deviation of  $\alpha$ from 1 makes $\gamma_{H}$ dependent on the exact frequency of measurement.

To summarize, we find that in metallic Ag (and  Cu) nanowires  the  low frequency  resistance fluctuation is enhanced when the  diameter is reduced.  The  power spectral density of the fluctuation has a 1/f$^{\alpha}$ character as seen in metallic films and wires of larger dimension.  The fluctuation has a significant low-frequency component and $\alpha$ increases from the usual 1 as the diameter d is reduced. The reduction of the diameter d also leads to an enhancement of the normalized fluctuation $\frac{<\Delta R^2>}{R^2}$. Our results shows that there are new features in the  1/f noise as the size is reduced and they become more prominent when the diameter of the wires fall to 15nm. It will be important to investigate  the origin of the new behavior. This is needed because 1/f noise will play an important role and may even become a limiting factor in use of such nanometer size metal wires as interconnects.

\subsection{Acknowledgments} 
The authors would like to thank Dr. N. Ravishankar (MRC, IISc) for the TEM data. This work is  supported by DST, Govt. of India, CSIR, Govt. of India  and DIT, Govt. of India. A. Bid would like to thank CSIR  for a fellowship and A.Bora wants to thank UGC for a fellowship.

\end{document}